\documentclass[journal=jacsat,manuscript=article]{achemso}

\usepackage{chemformula} 
\usepackage[T1]{fontenc} 

\DeclareUnicodeCharacter{2212}{\ensuremath{-}}

\title{Ultralow lattice thermal transport and considerable wave-like phonon tunneling in chalcogenide perovskite BaZrS$_3$}

\author{Yu Wu}
\email{291320917@qq.com}
\affiliation{Yangtze Delta Region Institute (Huzhou), University of Electronic Science and Technology of China, Huzhou 313001, China}

\author{Ying Chen}
\affiliation{School of Information Science and Technology, Fudan University, Shanghai 200433, China}

\author{Qiaoqiao Li}
\affiliation{School of Physics, University of Electronic Science and Technology of China, Chengdu 610054, China}

\author{Kui Xue}
\affiliation{School of Physics, University of Electronic Science and Technology of China, Chengdu 610054, China}

\author{Hezhu Shao}
\affiliation{College of Electrical and Electronic Engineering, Wenzhou University, Wenzhou, 325035, China}

\author{Hao Zhang}
\email{zhangh@fudan.edu.cn}
\affiliation{School of Information Science and Technology, Fudan University, Shanghai 200433, China}
\affiliation{Yiwu Research Institute of Fudan University, Chengbei Road, Yiwu City, Zhejiang 322000, China}

\author{Liujiang Zhou}
\email{ljzhou@uestc.edu.cn}
\affiliation{Yangtze Delta Region Institute (Huzhou), University of Electronic Science and Technology of China, Huzhou 313001, China}
\affiliation{School of Physics, University of Electronic Science and Technology of China, Chengdu 610054, China}

\begin{document}

\date{\today}

\begin{abstract}
Chalcogenide perovskites provide a promising avenue for non-toxic, stable thermoelectric materials. Here, thermal transport and thermoelectric properties of BaZrS$_3$ as a typical orthorhombic perovskite are investigated. An extremely low lattice thermal conductivity $\kappa_L$ of 1.84 W/mK at 300 K is revealed for BaZrS$_3$, due to the softening effect of Ba atoms on the lattice and the strong anharmonicity caused by the twisted structure. We demonstrate that coherence contributions to $\kappa_L$, arising from wave-like phonon tunneling, leading to a 18 \% thermal transport contribution at 300 K. The increasing temperature softens the phonons, thus reducing the group velocity of materials and increasing the scattering phase space. However, it simultaneously reduces the anharmonicity, which is dominant in BaZrS$_3$ and ultimately improves the particle-like thermal transport. Further, by replacing S atom with Se and Ti-alloying strategy, $ZT$ value of BaZrS$_3$ is significantly increased from 0.58 to 0.91 at 500 K, making it an important candidate for thermoelectric applications.
\end{abstract}

\flushbottom
\maketitle

\thispagestyle{empty}

\section*{Introduction}

The search for materials with low thermal conductivity is of great significance in the field of thermoelectric conversion. Perovskite materials typically exhibit low lattice thermal conductivity due to weak bonding and strong anharmonicity\cite{PANDEY2022,Tong2023}. Among them, organic−inorganic halide perovskites show great promise, but suffer from poor thermal and moisture stability\cite{Swarnkar2019}, along with the toxicity of Pb\cite{Babayigit2016}. The rise of halide perovskites encouraged researchers to explore oxide and chalcogenide perovskites in the electronic applications. Oxygen atoms have high electronegativity, leading to the large electronegativity difference along the B−X bond in ABX$_3$, which may enhance the ionization and electron localization. Hence, oxide perovskites usually have a wide band gap and poor electrical transport performance which is unfavorable to the thermoelectric properties of the material\cite{Swarnkar2019}.

The lower electronegativity difference of the B−X bond in chalcogenide perovskites makes them potential semiconductors for solar cell and other electronic applications. BaZrS$_3$ is a representative chalcogenide perovskite. It is a semiconductor with direct band gap of $\sim1.7-1.85\;\rm{eV}$ and shows strong absorption in the visible spectrum\cite{Meng2016,Swarnkar2019}. The band gap of BaZrS$_3$ can be easily tuned under pressure or by atomic substitution\cite{Gross2017,Niu2016}. BaZrS$_3$ is also shown to be stable against oxidation and moisture\cite{Perera2016}. Different from the ideal cubic perovskite, BaZrS$_3$ is isostructural with GdFeO$_3$ and adopts the twisted perovskite structure with orthorhombic phase\cite{Niu2019}. Studies have shown that the reduction of symmetry can significantly reduce the phonon degeneracy and enhance the avoid-crossing behavior of phonon branches, which result in an increase in the phonon−phonon scattering phase space to limit lattice thermal transport\cite{Liu2022,Liu2022a,Wu2023}. Using T=0 K force constants, the lattice thermal conductivity of BaZrS$_3$ at 300 K is calculated to be as low as 1.16 W/mK base on the iterative solution of Boltzmann Transport Equation (BTE)\cite{Osei-Agyemang2019}. However, in this scheme, the temperature effect only affects the Bose-Einstein distribution of phonons, and its effects on phonon dispersion, phonon-phonon interaction and electron-phonon interaction are completely ignored, which may lead to misestimation of thermal conductivity\cite{Shao2023}. In addition, the traditional particle-like treatment of phonons may not be sufficient for describing thermal transport properties in the complex BaZrS$_3$ structure. Sun et al. found large wave-like phonon tunneling effect through elastic and inelastic scattering measurements in hexagonal chalcogenide perovskite BaTiS$_3$\cite{Sun2020}. We expect that this phenomenon will also occur in orthorhombic BaZrS$_3$ and contribute to the total lattice thermal conductivity. Without considering the heat loss from lattice thermal transport, Osei-Agyemang et al. obtained an upper limit of 1.0 for the $ZT$ value of BaZrS$_3$ at 300 K\cite{OseiAgyemang2021}, indicating its potential application in the thermoelectric field. The accurate lattice thermal conductivity considering temperature and wave-like phonon tunneling effects is necessary for the further reliable estimation of $ZT$ value.

Here, we unveil the lattice thermal transport and thermoelectric properties of BaZrS$_3$ by means of first principles calculations. The conventional  particle-like behavior and wave-like behavior of phonons are all taken into account in lattice thermal transport calculations. We show that the temperature effect can reduce the phonon group velocity, increase the phase space of phonon scattering and reduce the anharmonicity of  BaZrS$_3$. Their influence on the particle-like thermal conductivity $\kappa_p$ is competitive with each other. Because the anharmonicity is more sensitive to temperature, the conventional fixed force constants method underpredicts the $\kappa_p$. We find significant contributions from wave-like transport channels in BaZrS$_3$ due to the closely spaced phonon branches. Further, we apply the lattice thermal conductivity including wave-like thermal transport to thermoelectric evaluation for the first time. By atom substitution method, we modulate the thermoelectric performance of BaZrS$_3$ and find that it is significantly improved after replacing S atom with Se or Ti-alloying, indicating their potential application in the thermoelectric field.

\section*{Results and Discussion}

\begin{figure*}[ht!]
\centering
\includegraphics[width=0.8\linewidth]{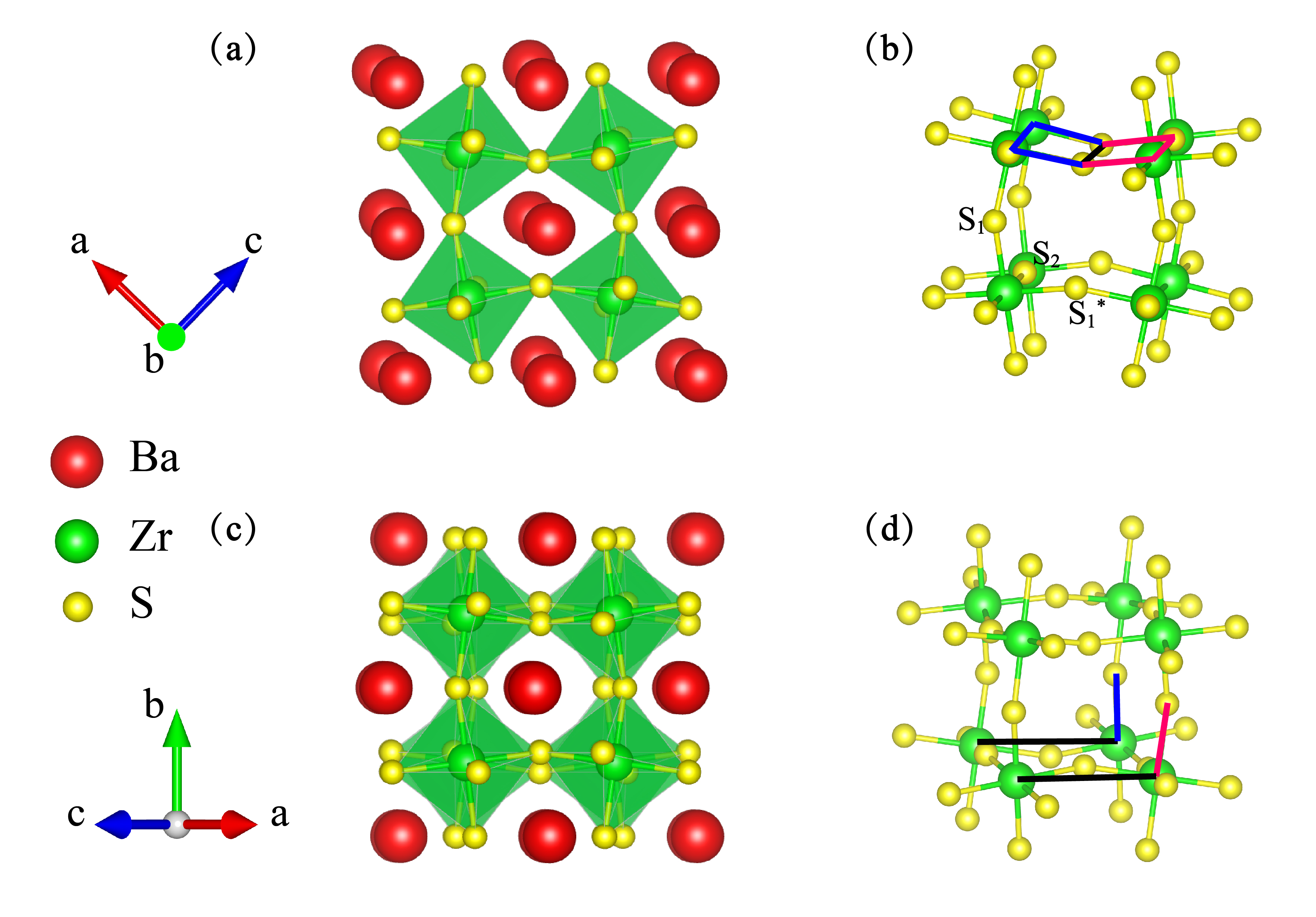}
\caption{(a, b) Crystal structure of BaZrS$_3$ viewed along $b$ axis with adjacent octahedral layers rotating in-phase. (c, d)  Crystal structure of BaZrS$_3$ viewed along [101] direction with adjacent octahedral layers rotating out-of-phase.}
\label{Fig1}
\end{figure*}

The configurations of BaZrS$_3$ are shown in Fig~\ref{Fig1}. BaZrS$_3$ has a twisted perovskite structure with a space group of $Pnma$. The framework of BaZrS$_3$ is composed of corner-shared ZrS$_6$ octahedra with Zr atom at the center and S atoms at the corners. The Ba atoms are located in the hollow within the perovskite framework. The optimized lattice parameters are listed in Table~\ref{tab1} which agree well with previous results\cite{OseiAgyemang2021,Niu2019}. The twisted structures are usually accompanied by small group velocities, large phase space and lattice anharmonicity\cite{Liu2022}, which will together limit the thermal transport of materials. From different directions, the ZrS$_6$ octahedra in BaZrS$_3$ have different distortion forms. Along the $b$ axis, the adjacent octahedral layers have in-phase rotation and are completely overlapping, as shown in Fig~\ref{Fig1}(a). The distortion along [101] direction is out-of-phase with the adjacent octahedral layers rotating in opposite directions, as shown in Fig~\ref{Fig1}(c). For in-phase distortion, the twisted angle is the half of the difference between $180^{\circ}$ and the dihedral angle of faces labeled in Fig~\ref{Fig1}(b). For out-of-phase distortion, the twisted angle is the half of two Zr-Zr-S angles labeled in Fig~\ref{Fig1}(d). The extracted in-phase and out-of-phase twisted angles are $9.07^{\circ}$ and $6.98^{\circ}$, respectively, close to the results from the experiment of $9.00^{\circ}$ and $7.03^{\circ}$\cite{Niu2019}. There are two kinds of S atoms in each unit cell of BaZrS$_3$, denoted as S$_1$ and S$_2$, respectively. There are two different Zr−S$_1$ bonds, i.e. Zr−S$_1$ and Zr−S$_1^*$ with the lengths of $2.541\;\rm{\AA}$ and $2.530\;\rm{\AA}$. The bond length between Zr and S$_2$ is relatively short, with a value of $2.524\;\rm{\AA}$. The bonding between Ba and S plays an important role in the lattice distortion. In BaZrS$_3$, Ba atoms distort eight S atoms from the neighboring eight adjacent ZrS$_6$ octahedra, with Ba-S bond lengths ranging from $3.160$ to $4.260\;\rm{\AA}$. The complicated twisted angles and bond lengths indicate a poor phonon transport properties of BaZrS$_3$. Moreover, the second-order interaction-force constants (2nd-order IFCs) of atomic pairs are extracted to reflect the bonding strength as shown in Fig S1. The greater the IFC of a atomic pair, the stronger the bonding strength. The Zr-S$_2$ bonding is the strongest among these atomic pairs, with the 2nd-order IFCs of $2.24\;\rm{eV\AA^{-2}}$. The Ba-S bonding is much weaker than Zr-S$_2$ bonding but comparable to Zr-S$_1$ and Zr-S$_1^*$. The difference of 2nd-order IFCs between Zr-S and Ba-S pairs is smaller than that between Sn-S and Ba-S pairs in BaSnS$_2$\cite{Li2021}. This is because compared with Sn atoms, the electronegativity of Zr atoms is closer to that of Ba atoms. Hence, the Ba atoms in BaZrS$_3$ will have a more significant effect on the distortion of the skeleton than in BaSnS$_2$. The bonding between Ba atoms and the Zr-S framework would significantly soften the lattice and limit the phonon transport.


\begin{table*}
\centering
\caption{The calculated lattice parameters (\AA), bong length (\AA) and band gap (eV) of BaZrS$_3$, compared with other calculated and experimental results.}
\setlength{\tabcolsep}{3mm}{
\begin{tabular}{cccccccc}
\hline
 & $a$ & $b$ & $c$ & $\rm{Zr-S_1}$ &  $\rm{Zr-S_1^*}$ & $\rm{Zr-S_2}$ & $E_g$ (eV)\\
\hline
 $this\,work$  & 7.084 & 9.944 & 6.960 & 2.541 & 2.530 & 2.524 & 1.79\\

 $Cal.$  & 7.066\cite{OseiAgyemang2021} & 9.923\cite{OseiAgyemang2021} & 6.972\cite{OseiAgyemang2021} & /  & /  &  / & 1.79\cite{OseiAgyemang2021}\\

 $Exp.$  & 7.056\cite{Niu2019} & 9.962\cite{Niu2019} & 6.996\cite{Niu2019} & 2.539\cite{Niu2019} & 2.532\cite{Niu2019} & 2.528\cite{Niu2019} & 1.70-1.85\cite{Niu2016,Perera2016,Meng2016}\\
\hline
\end{tabular}
}
\label{tab1}
\end{table*}

\begin{figure*}[ht!]
\centering
\includegraphics[width=1\linewidth]{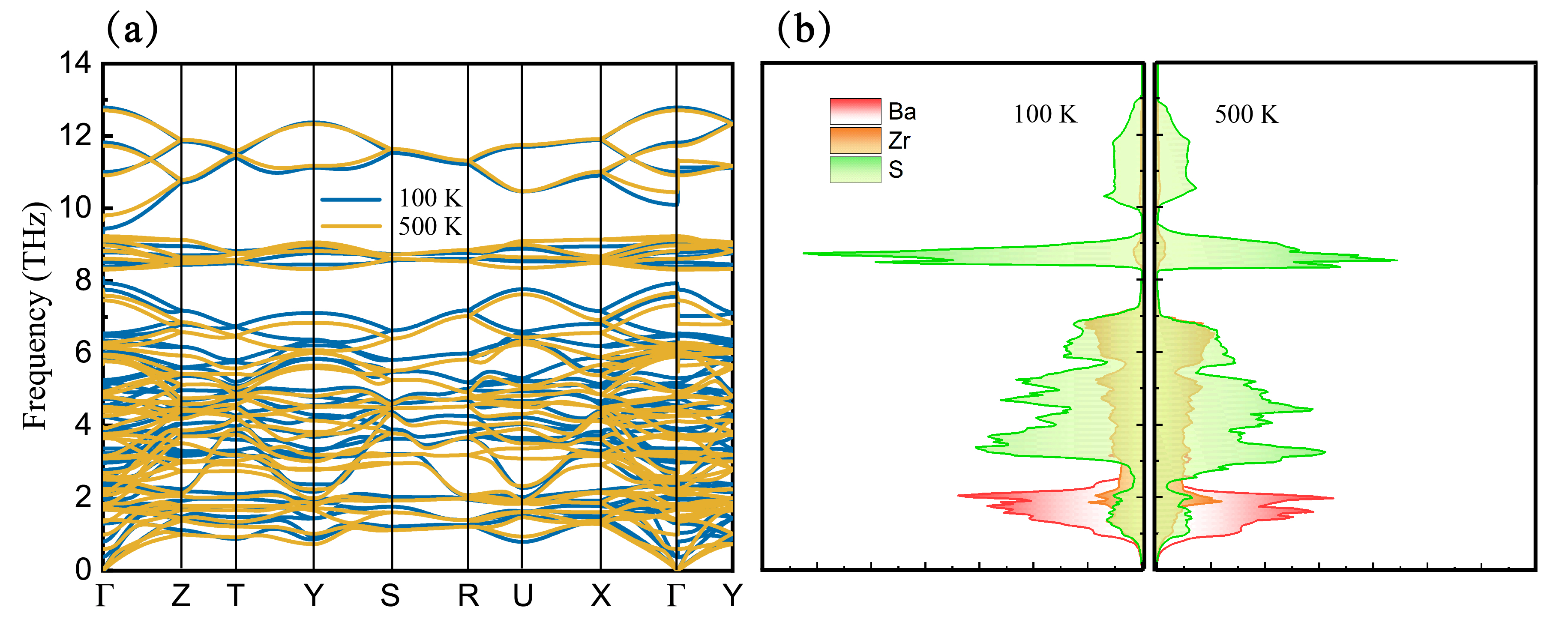}
\caption{ (a) Calculated temperature-dependent phonon dispersions of BaZrS$_3$ at 100 K and 500 K. (b) Atom-resolved phonon density of states of BaZrS$_3$ at 100 K and 500 K.}
\label{phband_dos}
\end{figure*}

The phonon dispersion and atom-resolved density of states for BaZrS$_3$ at 100 K and 500 K are presented in Fig~\ref{phband_dos}. In BaZrS$_3$, most phonon branches are much flatter than those in ZrS$_3$ and the maximum frequency of acoustic phonon modes for BaZrS$_3$ is 1.5 THz at 100 K much smaller than that of 3.9 THz in ZrS$_3$\cite{Wang2020a}. It indicates that the phonon modes in BaZrS$_3$ have lower group velocity due to the softening effect of Ba atoms on the lattice. Moreover, the optical phonon branches in the range of 2-4 THz show larger slopes than acoustic phonon branches, which indicates higher phonon group velocities. The optical phonon branches are close to each other. On one hand, it will increase the phonon−phonon scattering phase space by increasing the scattering channels. On the other hand, it gives optical phonons the opportunity to couple with each other and benefits to the wave-like tunneling transport. From the atom-resolved density of states, it can be seen that the acoustic and low frequency optical branches are mainly contributed by Ba atoms, while the high frequency optical branches are mainly contributed by Zr and S atoms. With the increase of temperature, the phonon frequencies decrease overall and the density of states tend to be compressed.

In the framework of the Wigner formalism\cite{Moyal1949}, the lattice thermal conductivity is generalized to an expression including both the populations' and coherences' contributions\cite{Simoncelli2019}:

\begin{equation}
\kappa_L^{\alpha \beta}=\kappa_{\mathrm{p}}^{\alpha \beta}+\kappa_{\mathrm{c}}^{\alpha \beta}
\end{equation}

where $\kappa_{\mathrm{p}}^{\alpha \beta}$  is related to the particle-like propagation of phonon wavepackets discussed by Peierls's semiclassical picture, written as

\begin{equation}
\kappa_{\mathrm{p}}^{\alpha \beta}=\frac{1}{V N} \sum_{\mathrm{q}\mathrm{s}} C_{\mathrm{q}}^s v_{\mathrm{q}, \alpha}^s v_{\mathrm{q}, \beta}^s \tau_{\mathrm{q}}^s
\end{equation}

$\kappa_{\mathrm{c}}^{\alpha \beta}$ is associated to the wave-like tunnelling and loss of coherence between different phonon branches $s$ and $s^{\prime}$, written as

\begin{equation}
\begin{aligned}
\kappa_{\mathrm{c}}^{\alpha \beta}= & \frac{\hbar^2}{k_B T^2 V N} \sum_{\mathrm{q}} \sum_{s \neq s^{\prime}} \frac{\omega_{\mathrm{q}}^s+\omega_{\mathrm{q}}^{s^{\prime}}}{2} v_{\mathrm{q}, \alpha}^{s, s^{\prime}} v_{\mathrm{q}, \beta}^{s, s^{\prime}} \\
& \times \frac{\omega_{\mathrm{q}}^s n_{\mathrm{q}}^s\left(n_{\mathrm{q}}^s+1\right)+\omega_{\mathrm{q}}^{s^{\prime}} n_{\mathrm{q}}^{s^{\prime}}\left(n_{\mathrm{q}}^{s^{\prime}}+1\right)}{4\left(\omega_{\mathrm{q}}^s-\omega_{\mathrm{q}}^{s^{\prime}}\right)^2+\left(\Gamma_{\mathrm{q}}^s+\Gamma_{\mathrm{q}}^{s^{\prime}}\right)^2} \times\left(\Gamma_{\mathrm{q}}^s+\Gamma_{\mathrm{q}}^{s^{\prime}}\right)
\end{aligned}
\end{equation}

In these expressions, $\alpha$ and $\beta$ are the Cartesian indices, $V$ is the the volume of the unit cell, $N$ is  the number of sampled phonons in the first Brillouin zone. $C_{\mathrm{q}}^s$, $v_{\mathrm{q}}^s$, $\tau_{\mathrm{q}}^s$, $\omega_{\mathrm{q}}^s$, $v_{\mathrm{q}}^{s, s^{\prime}}$, $\Gamma_{\mathrm{q}}^s$ are the heat capacity, the group velocity, the lifetime, the phonon frequency, the interband velocity matrix element and the scattering rate of a phonon mode with wavevector $\mathrm{q}$. $n_{\mathrm{q}}^s$ is the equilibrium Bose–Einstein distribution. 

\begin{figure*}[ht!]
\centering
\includegraphics[width=1\linewidth]{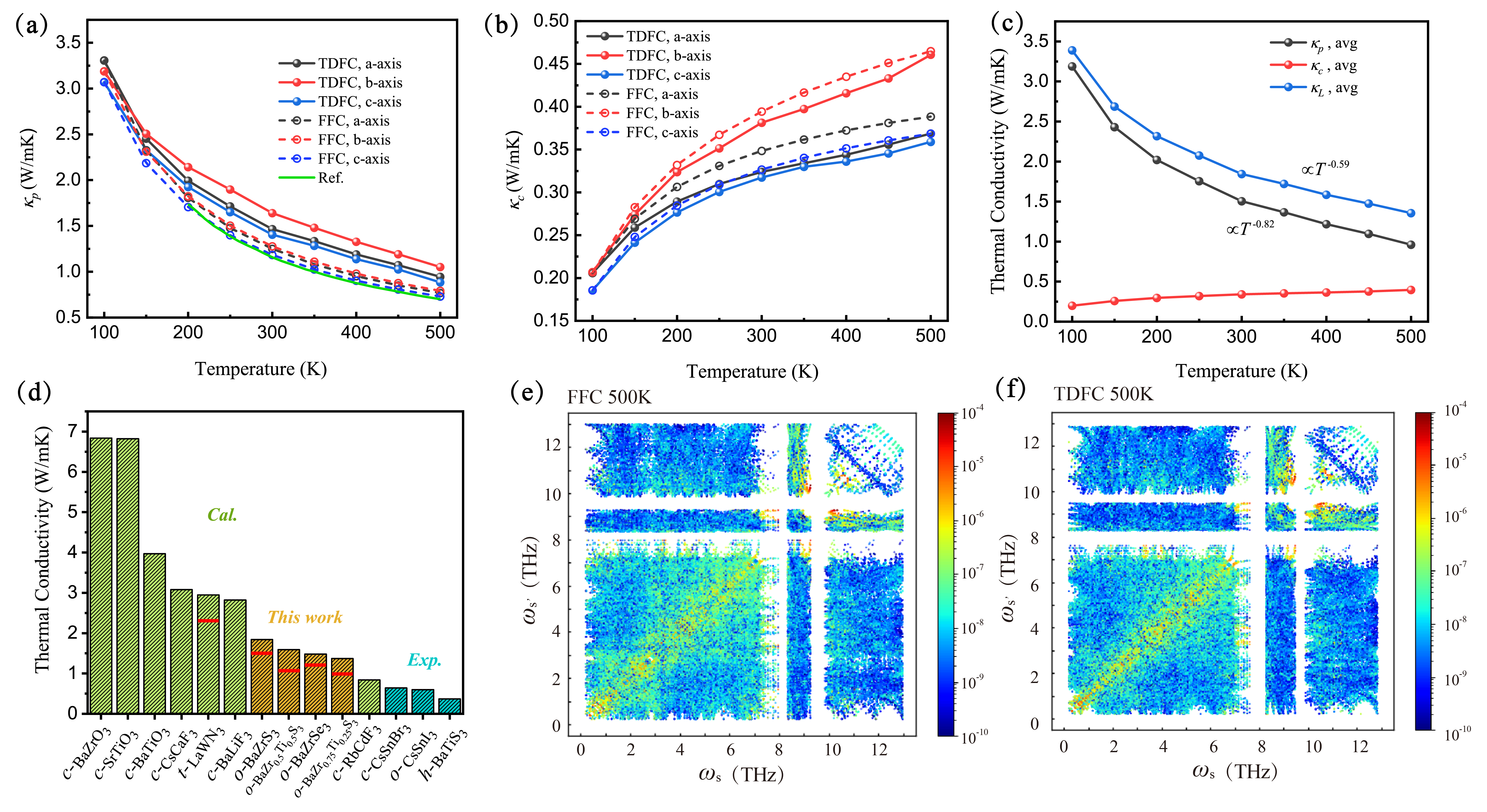}
\caption{ Calculated temperature-dependent (a) $\kappa_{\mathrm{p}}$ and (b) $\kappa_{\mathrm{c}}$ using FFC and TDFC considering three-phonon interaction. The result from reference is also shown with green line\cite{Osei-Agyemang2019}. (c) The temperature-dependent average $\kappa_{\mathrm{p}}$, $\kappa_{\mathrm{c}}$ and $\kappa_L$. (d) $\kappa_L$ from other literature of some other inorganic perovskites at 300 K.\cite{Zhao2021,Tong2023,Xie2020,Sun2020} $\kappa_L$ of LaWN$_3$, BaZrS$_3$, BaZrSe$_3$, BaZr$_{0.75}$Ti$_{0.25}$S$_3$ and BaZr$_{0.5}$Ti$_{0.5}$S$_3$ contains $\kappa_{\mathrm{c}}$ contribution, which is represented by the part above the solid red line. The prefixes $c-$, $o-$, $t-$ and $h-$ represent cubic, orthorhombic, trigonal and hexagonal, respectively. The resolved $\kappa_{\mathrm{c}}$ with various pairs of phonon frequencies ($\omega^s$ and $\omega^{s^{\prime}}$) at 500 K using (e) FFC and (f) TDFC methods.}
\label{kappa}
\end{figure*}

Fig~\ref{kappa}(a) shows the anisotropic $\kappa_{\mathrm{p}}$ of BaZrS$_3$ using the fixed force fonstants (FFC) at 100 K and temperature dependent force constants (TDFC) considering three-phonon scattering. The calculated $\kappa_{\mathrm{p}}$ is 1.25 W/mK, 1.28 W/mK and 1.18 W/mK along the $a-$, $b-$ and $c-$axis at 300 K using FFC method, which are close to the previous results of Eric $et\,al$.\cite{Osei-Agyemang2019}. The decrease of thermal conductivity with temperature becomes slow when considering TDFC, making $\kappa_{\mathrm{p}}$ increase to 1.47 W/mK, 1.64 W/mK and 1.41 W/mK in the corresponding directions. This phenomenon has also been reported in UO$_2$ \cite{Yang2022}, but the opposite trend was found in LaWN$_3$ recently\cite{Tong2023}, which we will discuss later. As shown in Fig~\ref{kappa}(b), $\kappa_{\mathrm{c}}$ increases with the increasing temperature. The $\kappa_{\mathrm{c}}$ is largest along the $b$ direction, with 0.38 W/mK at 300 K using TDFC. It can be seen in Fig~\ref{phband_dos} that the phonon dispersion is denser and flatter along $\Gamma$-Z direction (related to $b$ direction), making it easy for two phonons coupling with each other. Fig~\ref{kappa}(c) shows the average thermal conductivity by calculating the arithmetic average along three axes. Since $\kappa_{\mathrm{p}}$ and $\kappa_{\mathrm{c}}$ show an opposite trend with temperature, the contribution of $\kappa_{\mathrm{c}}$ to $\kappa_L$ gradually increases with the increase of temperature. At 300 K, $\kappa_L$ is 1.84 W/mK and $\kappa_{\mathrm{c}}$ reaches 18\% of the $\kappa_L$. Moreover, $\kappa_{\mathrm{c}}$ makes the temperature dependence of thermal conductivity weaker, easing the decay from $T^{-0.82}$ to $T^{-0.59}$. Fig~\ref{kappa}(c) shows the thermal conductivity of BaZrS$_3$ compared with some other inorganic perovskites. Our results show that the low thermal conductivity of BaZrS$_3$ has a competitive advantage, indicating its potential application in the field of thermoelectric. The lattice thermal conductivity can be reduced by reducing the symmetry of the material or by introducing heavy atoms\cite{Li2018abc}. On the basis of BaZrS$_3$ structure, by atomic substitution method, three materials of BaZrSe$_3$, BaZr$_{0.75}$Ti$_{0.25}$S$_3$ and BaZr$_{0.5}$Ti$_{0.5}$S$_3$ are obtained and their lattice thermal conductivity are calculated. Compared with BaZrS$_3$, they all have lower lattice thermal conductivity. For example, by replacing the S atom with the heavier Se, the lattice thermal conductivity of the obtained BaZrSe$_3$ can decrease to 1.48 W/mK.

In Fig~\ref{kappa}(e, f), we depicted the $\kappa_{\mathrm{c}}$ terms resolved by various pairs of phonon frequencies at 500 K. The phonon pairs around the diagonal ($\omega^s=\omega^{s^{\prime}}$) have a larger contribution to $\kappa_{\mathrm{c}}$. Compared with FFC method, the $\kappa_{\mathrm{c}}$ is more concentrated when considering TDFC due to the softening of phonon dispersion with the increase of temperature.

\begin{figure*}[ht!]
\centering
\includegraphics[width=0.9\linewidth]{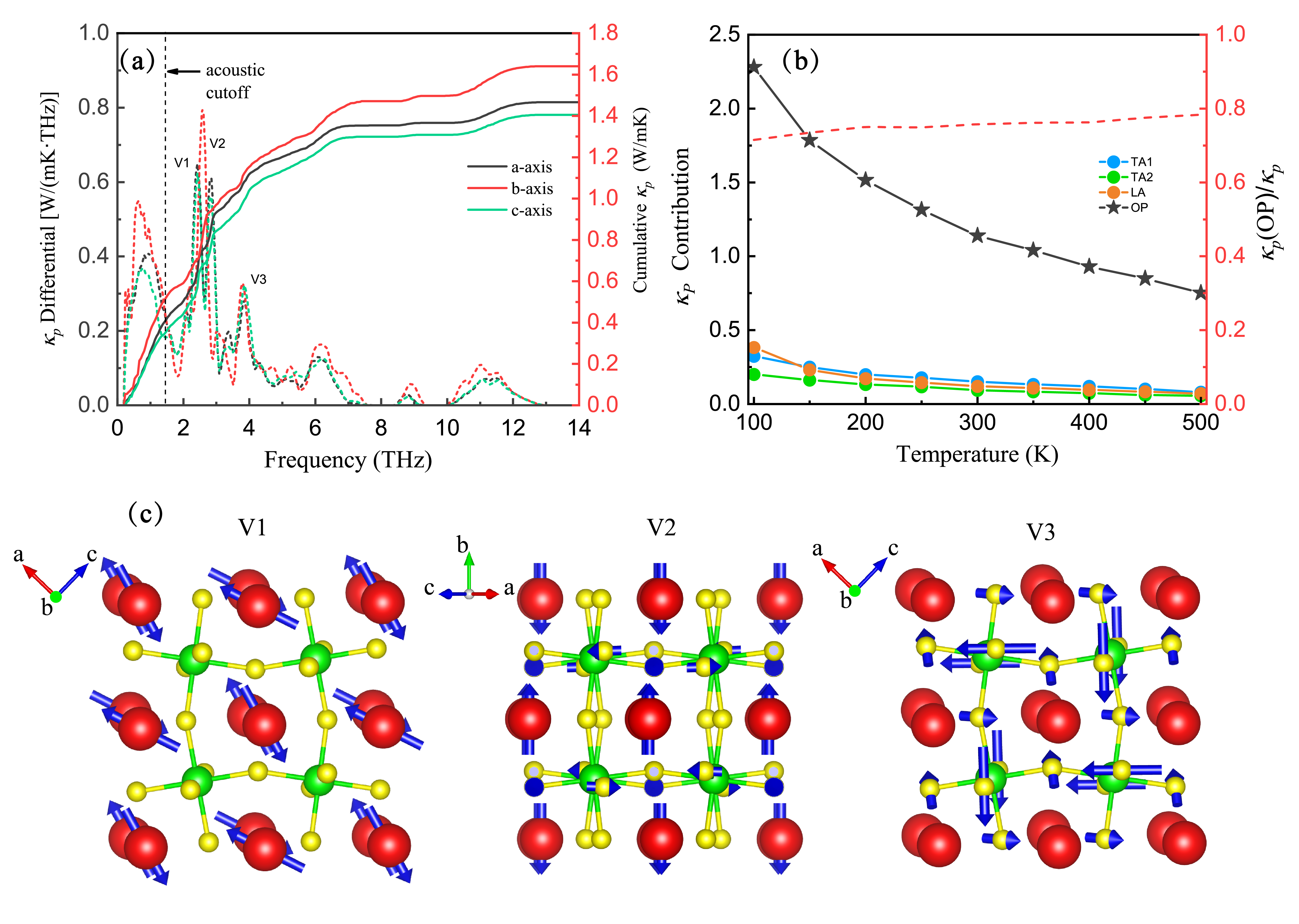}
\caption{ (a) Calculated cumulative $\kappa_{\mathrm{p}}$ and differential $\kappa_{\mathrm{p}}$ as a function of phonon frequencies. (b) Temperature
dependence of $\kappa_{\mathrm{p}}$ contributed from three acoustic phonon branches (TA1, TA2 and LA) and all the optical branches (OP). (c) Atomic vibrations of $\Gamma$-point phonon modes corresponding to the peaks with the most $\kappa_{\mathrm{p}}$ contribution.}
\label{diff_kappa}
\end{figure*}

To investigate the contribution of phonons with different frequencies to $\kappa_{\mathrm{p}}$, Fig~\ref{diff_kappa}(a) shows the frequency dependent cumulative $\kappa_{\mathrm{p}}$ and the differential $\kappa_{\mathrm{p}}$ at 300 K. The most significant increase in thermal conductivity is observed in two regions with the frequency ranges of 0-2 THz and 2-4 THz. The former mainly contains acoustic phonons. The latter related to low-frequency optical phonons which contributes 42\% to the total $\kappa_{\mathrm{p}}$ along the $a$-axis. As seen in Fig~\ref{diff_kappa}(b), the contribution of optical phonons to the average $\kappa_{\mathrm{p}}$ maintains around 75\% in the temperature ranging from 100 K to 500 K, close to 68\% in BaSnS$_2$ and 71\% in SnS\cite{Li2021}. Atomic vibrations of $\Gamma$-point phonon modes corresponding to V1, V2 and V3 in Fig~\ref{diff_kappa}(a) are shown in Fig~\ref{diff_kappa}(c). For the V1 mode, where Ba vibrations dominate in the $a-c$ plane, the neighboring Ba atoms along $b$ and [101] direction both vibrate in opposite direction. The V2 mode is also mainly contributed by the vibration of Ba atoms, but vibrating along $b$ direction. The neighboring Ba atoms along $b$ direction vibrate in opposite direction but in the same direction along [101] direction. The V3 mode is mainly contributed by S$_2$ vibrations. The S$_2$ atoms in one octahedra vibrate in the same direction but vibrate in a vertical direction in the adjacent octahedra. The vibrations along different directions lead to significant deformation for the lattice, which will further bring huge anharmonicity.

\begin{figure*}[ht!]
\centering
\includegraphics[width=1\linewidth]{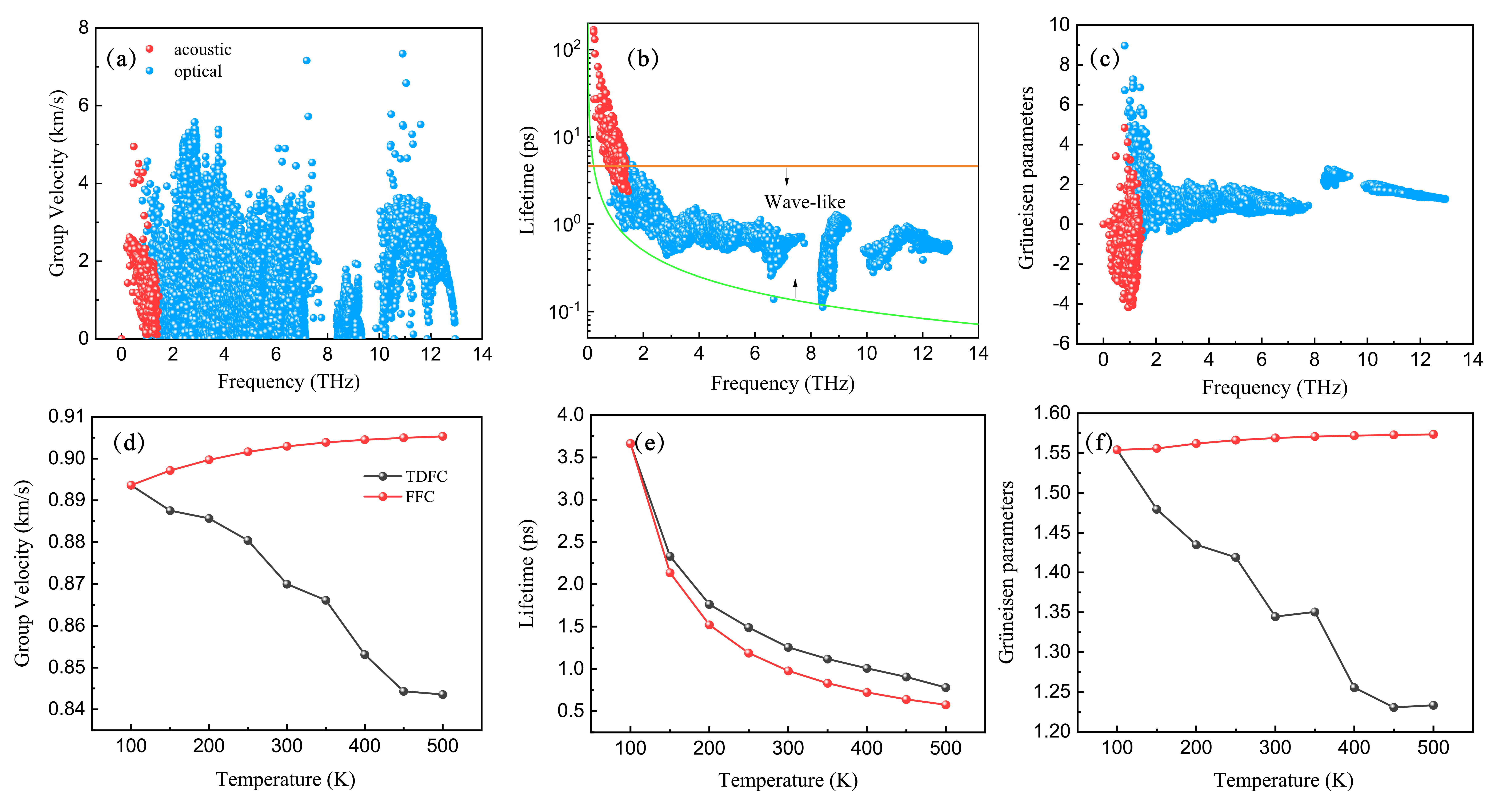}
\caption{Mode-resolved (a) group velocity (b) lifetime, and (c) Gr{\"u}neisen parameter versus phonon frequency in BaZrS$_3$ at 300 K using TDFC. (d-f) Temperature dependent average group velocity, lifetime, and Gr{\"u}neisen parameter using TDFC and FFC.}
\label{Vg}
\end{figure*}

To further understand the underlying mechanisms of the low lattice thermal conductivities and the dominant contribution by optical phonons for BaZrS$_3$, we also calculate the corresponding group velocity $V_g$, phonon lifetime $\tau$, and  Gr{\"u}neisen parameter $\gamma$. As shown in Fig~\ref{Vg}(a), most of the phonons have group velocity less than 2 km/s due to the flat band properties seen in Fig~\ref{phband_dos}(a). Moreover, the group velocity of optical phonons is generally higher than that of acoustic phonons, especially for optical phonons in the 2-4 THz frequency range. These phonons with large group velocity will mainly contribute to the thermal conductivity of the material. The average group velocity can be defined as

\begin{equation}
\bar{V_g}=\sqrt{\frac{\sum_{\mathrm{q} s} C_{{\mathrm{q} s}} v_{\mathrm{q} s}^2 \tau_{\mathrm{q} s} N_{\mathrm{q} s}}{\sum_{\mathrm{q} s} C_{{\mathrm{q} s}} \tau_{\mathrm{q} s} N_{\mathrm{q} s}}},
\end{equation}

where $N$ is the symmetry degeneracy of corresponding mode $\mathrm{q} s$. The calculated average group velocity is 0.87 km/s for BaZrS$_3$, comparable to that of low thermal conductivity material BaSnS$_2$ with 0.8 km/s\cite{Li2021}. Fig~\ref{Vg}(b) shows the mode resolved phonon lifetime. The lifetime of acoustic phonons is in the range of 1-100 ps, while is 0.1-10 ps for the optical phonons. Shorter phonon lifetimes indicate greater scattering and weaker contribution to thermal transport. Hence, optical phonons dominating thermal transport in BaZrS$_3$ is mainly due to their large group velocity. Fig S2 also shows the scattering rates contributed by the interaction of three and four phonons. The four-phonon scattering rate is more than two orders of magnitude smaller than the three-phonon scattering rate in a wide frequency range, indicating that it is reasonable to consider only the three-phonon interaction. The average lifetime can be written as

\begin{equation}
\bar{\tau}=\frac{\sum_{\mathrm{q} s} C_{\mathrm{q} s} v_{\mathrm{q} s}^2 \tau_{\mathrm{q} s}}{\sum_{\mathrm{q} n} C_{\mathrm{q} s} v_{\mathrm{q} s}^2}.
\end{equation}

The calculated $\bar{\tau}$ is 1.26 ps for BaZrS$_3$, lower than that in BaSnS$_2$ with 2.37 ps\cite{Li2021}, indicating the potential lattice anharmonicity in BaZrS$_3$. Besides, we identify the signature of a wave-like tunneling behavior in the heat carrying phonons. The horizontal lines in Fig~\ref{Vg}(b) is the Wigner limit $\tau_{Wigner}$ reflecting the average phonon interband spacing, i.e., $\tau_{Wigner}= N_s/\omega_{\max}$, where $N_s$ is the total number of phonon bands and $\omega_{\max}$ is the maximum phonon frequency\cite{Simoncelli2022}. The green curve gives the phonon lifetime in the Ioffe-Regel limit ($\tau_{Ioffe-Regel}= 1/\omega$)\cite{Allen1993}.  The majority of phonon lifetime are in the $\tau_{Ioffe-Regel}<\tau<\tau_{Wigner}$ range resulting in large $\kappa_{\mathrm{c}}$ contributions. The Gr{\"u}neisen parameter $\gamma$ reflects the anharmonic property of the lattice determined by the change of phonon frequencies with respect to the change of unit-cell volume. The higher the absolute value of $\gamma$, the stronger the lattice anharmonicity. As seen in Fig~\ref{Vg}(c), 
acoustic phonons and low-frequency optical phonons have high $\gamma$ range from -4 to 9. The average Gr{\"u}neisen parameter is defined as

\begin{equation}
\bar{\gamma}=\frac{\sum_{\mathrm{q} s}\left|\gamma_{\mathrm{q} s}\right| C_{{\mathrm{q} s}}}{\sum_{\mathrm{q} s} C_{{\mathrm{q} s}}}
\end{equation}

The calculated $\bar{\gamma}$ is 1.34 for BaZrS$_3$ comparable to some materials with strong anharmonicity such as PbTe ($\gamma\sim1.40$)\cite{Morelli2008} and BiCuSeO ($\gamma\sim1.50$)\cite{Pei2013}.

To understant where differences in thermal transport originate for TDFC and FFC calculations, we compare the $\bar{V_g}$, $\bar{\tau}$ and $\bar{\gamma}$ at different temperatures derived from the two methods. As shown in Fig~\ref{Vg}(d), the $\bar{V_g}$ using FFC is around 0.9 km/s but gradually decreases as the temperature rises when using TDFC. This result is consistent with that the phonon dispersion is softened with the increase of temperature as shown in Fig~\ref{phband_dos}(a). The change rate of $\bar{V_g}$ is -3.7\% after using TDFC at 300 K. This change in $\bar{V_g}$ will limit the thermal transport performance of the material. As the temperature increases, the $\bar{\tau}$ decreases gradually for both methods, but at each temperature, the results obtained by TDFC method are larger. The change rate of $\bar{\tau}$ is 28.4\% after using TDFC at 300 K, which will improve the thermal conductivity. Hence, the differences in thermal transport using TDFC and FFC calculations lie in the change of phonon lifetime. In fact, the lifetime of phonons are determined by the strength of anharmonic scatterings and the scattering possibilities. The former is related to the anharmonicity and can be characterized by $\gamma$. The latter is determined by the three-phonon scatterings which include emission and absorption processes and can be characterized by the phase space $P_3$. Fig S3 shows $P_3$ at 300 K using FFC and TDFC methods. It can be seen that the temperature effect only causes a weak enhancement of the scattering channel in the low frequency region. As shown in Fig~\ref{Vg}(f), the $\bar{\gamma}$ using FFC method remains almost constant ($\sim1.57$), while generally decreases as the temperature rises using TDFC. It indicates that the temperature effect can lower the anharmonicity of BaZrS$_3$, which further increases the phonon lifetime at a specific temperature and thus delays the trend of thermal conductivity decreasing with increasing temperature shown in Fig~\ref{kappa}(a).

To precisely evaluate the electronic transport properties of BaZrS$_3$, ionized impurity (IMP) scattering, acoustic deformation-potential (ADP) scattering and polar optical phonon (POP) scattering are taken into considerations. Fig S4(a, b) show the electron and hole mobility decomposed by various scattering mechanisms with carrier concentration $n=10^{19}\;\rm{cm^{-3}}$. BaZrS$_3$ is a polar perovskite material, in which there is a large dielectric electron-phonon interaction, yet the kinetic energy of the electron is sufficient to keep the polaron from collapsing into a localized state\cite{Frost2017}. In this material the polar-optical phonon mode scattering will dominate and thus limit mobility. In BaZrS$_3$, the calculated electron and hole mobility are $37.9\;\rm{cm^2/V\cdot s}$ and $4.3\;\rm{cm^2/V\cdot s}$ at 300 K, respectively, close to that of CH$_3$NH$_3$PbI$_3$ $\sim30\;\rm{cm^2/V\cdot s}$\cite{Herz2017}. Fig S4(c, d) show the electron and hole scattering rates for IMP, ADP and POP scattering. The POP scattering rates of most electrons and holes are $\sim10^{14}\;\rm{s^{-1}}$, which is 1-2 orders of magnitude higher than the scattering rates of IMP and ADP. Our previous work suggests that  Se-alloying strategy can weaken the electron−phonon coupling and prolong the nonradiative electron−hole recombination lifetime in BaZrS$_3$ while the opposite is true when using the Ti-alloying strategy\cite{Li2022}. The calculated electron and hole mobility are $49.5\;\rm{cm^2/V\cdot s}$ and $17.0\;\rm{cm^2/V\cdot s}$ at 300 K for BaZrSe$_3$, which confirms the superior electrical properties of BaZrSe$_3$ over BaZrS$_3$. The electron mobility of  BaZr$_{0.75}$Ti$_{0.25}$S$_3$ and BaZr$_{0.5}$Ti$_{0.5}$S$_3$ decrease by about half compared with BaZrS$_3$ as shown in Fig S5(a).

\begin{figure*}[ht!]
\centering
\includegraphics[width=0.9\linewidth]{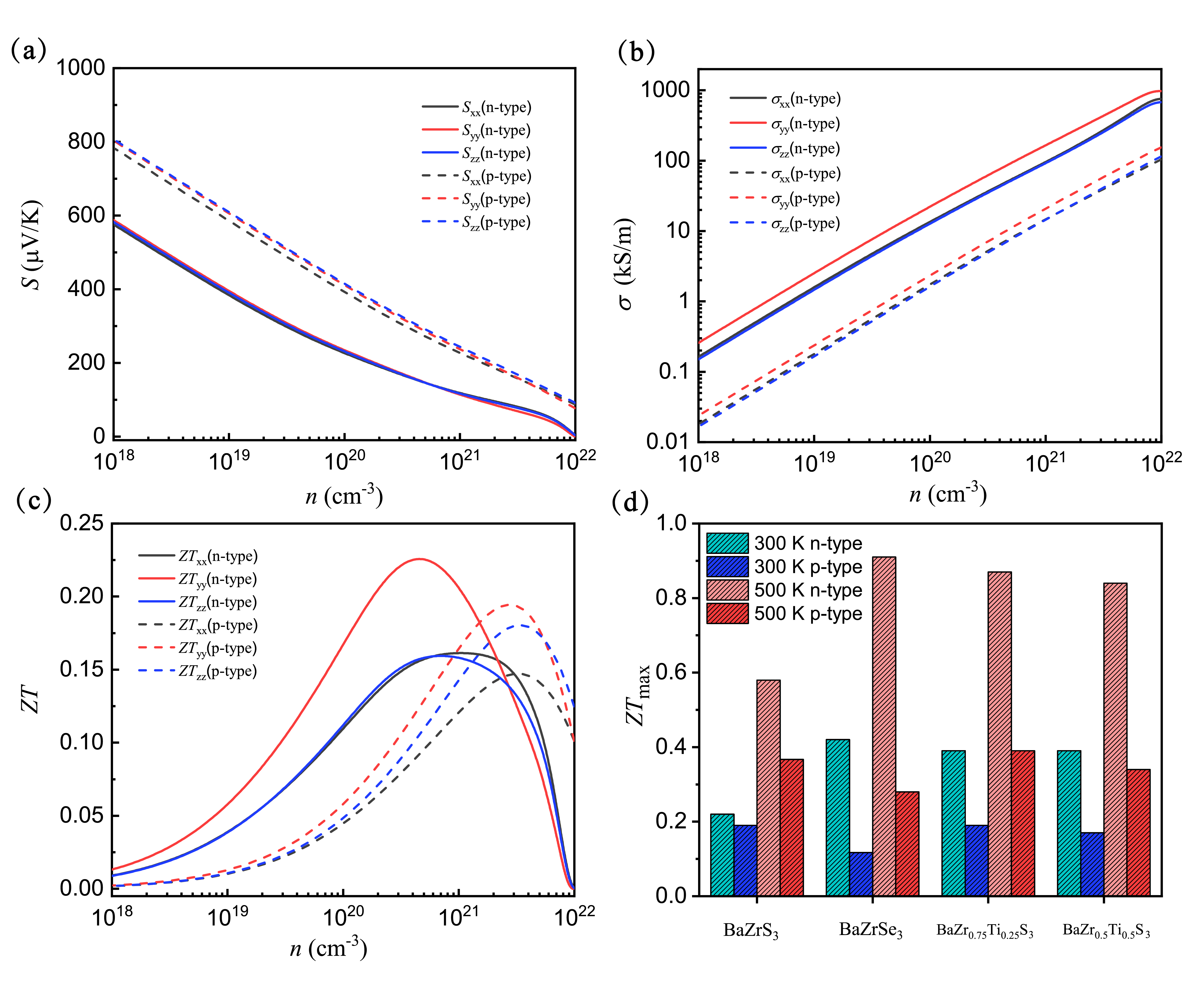}
\caption{The thermoelectric properties for BaZrS$_3$. (a) seebeck coefficients, (b) electronic conductivities and (c) $ZT$ values vesus carrier concentration at 300 K. (d) The maximum $ZT$ at different temperatures.}
\label{TE}
\end{figure*}

The thermoelectric performance of a material is quantified by a dimensionless figure of merit $ZT$ ($ZT=S^2\sigma T/(\kappa_{e}+\kappa_L)$), where $S$ represents the seebeck coefficient, $\sigma$ is the electron conductivity, and $\kappa_{e/L}$ is the electronic/lattice thermal conductivity. The thermoelectric parameters of seebeck coefficients $S$ and electronic conductivities $\sigma$ varying with carrier concentration $n$ at 300 K are shown in Fig~\ref{TE}. The band- and momentum-resolved relaxation time are considered in the calculation including ADP and POP scattering. The $S$ is almost isotropic and the average value is predicted to be $389\;\rm{\mu V/K}$ and $597\;\rm{\mu V/K}$ at $n=10^{19}\;\rm{cm^{-3}}$ for n-type and p-type system, respectively. The $S$ for BaZrS$_3$ is comparable to some traditional thermoelectric materials including Bi$_2$Te$_3$ ($215\;\rm{\mu V/K}$)\cite{Poudel2008}, PbTe ($185\;\rm{\mu V/K}$)\cite{Nielsen2012}, and SnSe ($510\;\rm{\mu V/K}$)\cite{Zhang2015}. The large $S$ for BaZrS$_3$ can be understood by the Mott relation\cite{Sun2015} that large seebeck coefficients tend to appear on the edge of the band with large DOS. Fig S6 shows the band structure and DOS of BaZrS$_3$. Both the valence band and the conduction band of BaZrS$_3$ have multi-valley features, leading large DOS on both sides of the Fermi level. $\sigma$ increases gradually with the increase of carrier concentration. At $n=10^{19}\;\rm{cm^{-3}}$, the average $\sigma=0.20\;\rm{kS/m}$ under p-type doping is  significantly lower than that of competing electronic materials. For n-type system, $\sigma$ has a higher magnitude of $\sigma=1.92\;\rm{kS/m}$ because of the lower carrier effective mass and scattering rate for electrons compared with holes. $\kappa_{e}$ can be calculated according to the Wiedemann-Franz law, i.e. $\kappa_e=L\sigma T$, where $L$ is the Lorenz number. The dimensionless figure of merit $ZT$ vesus carrier concentration at 300 K for BaZrS$_3$ are calculated and shown in Fig~\ref{TE}(c). A maximum value of $ZT=0.19$ at $n=2.6\times10^{21}\;\rm{cm^{-3}}$ and $ZT=0.23$ at $n=4.1\times10^{20}\;\rm{cm^{-3}}$ is realized for p-type and n-type system, respectively. As the temperature increases, the maximum $ZT$ increases further up to 0.58 at 500 K mainly due to the decrease of lattice thermal conductivity as shown in Fig~\ref{TE}(d). 

The band structure of BaZrSe$_3$, BaZr$_{0.75}$Ti$_{0.25}$S$_3$ and BaZr$_{0.5}$Ti$_{0.5}$S$_3$ are shown in Fig S7. Compared with BaZrS$_3$, their band gap decreases significantly and the multi-valley features are inherited, which are favorable for thermoelectric performance. Fig S5(b) shows the seebeck coefficients when $ZT$ reaches its maximum value. It can be seen that, compared with BaZrS$_3$, the electron seebeck coefficients are significantly increased for both BaZrSe$_3$, BaZr$_{0.75}$Ti$_{0.25}$S$_3$ and BaZr$_{0.5}$Ti$_{0.5}$S$_3$. Due to lower lattice thermal conductivity and better electrical transport performance, after replacing S element with Se element in BaZrS$_3$, the maximum $ZT$ of n-type BaZrSe$_3$ can reach 0.42 at 300 K, which is much higher than the inorganic perovskite CsPbI$_3$ (0.08) and comparable to the hybrid perovskite $\delta$-FAPbI$_3$ (0.45)\cite{Shi2023}. The $ZT$ can be further increased to 0.91 at 500 K.

\section*{Conclusion}

In conclusion, we investigate the lattice thermal transport and thermoelectric properties of BaZrS$_3$ through first-principles calculations. BaZrS$_3$ have low lattice thermal conductivity $\kappa_L$ with 1.84 W/mK at 300 K due to weak bonding and strong anharmonicity. The twisted structure and bonding between Ba atoms and the Zr-S framework significantly soften the lattice and lower the phonon group velocity. Because of the relatively high group velocity of optical phonons, they make a major contribution to thermal transport. We investigate how the temperature effect influence the thermal transport. The temperature effect mainly reduces the anharmonicity of the material and thus reduces the decrease of particle-like thermal conductivity with temperature. Our work also highlights the importance of wave-like thermal transport contributions to $\kappa_L$, especially at high temperatures. The polar optical phonons dominate the carrier scattering in BaZrS$_3$ and the electron mobility is $37.9\;\rm{cm^2/V\cdot s}$ at 300 K with $n=10^{19}\;\rm{cm^{-3}}$. BaZrS$_3$ has high seebeck coefficient due to multi-valley band structure. At 500 K, the maximum $ZT$ value is 0.58 for BaZrS$_3$ and can be further improved by replacing the S element with Se or Ti-alloying strategy, indicating excellent thermoelectric performance in the perovskite family. 

\section*{Numerical methods}

Calculations relating to electrical properties are implemented using the Vienna Ab Initio simulation package (VASP) based on density functional theory (DFT)\cite{Kresse1996}. The revised PBE-GGA for solids (PBEsol)\cite{Perdew2008} is evaluated as the exchange-correlation functional with a kinetic energy cutoff of $500\;\rm{eV}$. The energy convergence value between two consecutive steps is set as $10^{-5}\;$eV when optimizing atomic positions and the maximum Hellmann-Feynman (HF) force acting on each atom is $10^{-3}\;$eV/\r{A}. The reciprocal space is sampled by a grid of $7\times5\times7$ k points in the Brillouin zone. For the band gap calculations, the hybrid HSE06 functional\cite{Heyd2003} is used with the screening parameter setting as 0.25$\;$\r{A}$^{-1}$. The scattering rates and carrier mobilities are calculated using the AMSET package\cite{Ganose2021} considering ionized impurity (IMP) scattering, acoustic deformation-potential (ADP) scattering and polar optical (POP) scattering mechanism. The obtained band- and momentum-relevant scattering rates are then used as inputs of the BoltzTrap2 package\cite{Madsen2018} to evaluate the thermoelectric properties of the material. 

The calculations of $\kappa$ and other relevent parameters such as phonon relaxation time are carried out by the ShengBTE software\cite{shengbte2014} which operates based on the iterative scheme. The code used to calculate wave-like thermal conductivity has been integrated into ShengBTE software by us. The $\bf{q}$-mesh in the first irreducible Brillouin Zone is set be 13$\times$11$\times$13, the scale parameter for Gaussian smearing is set to be 0.1, and the supercells of 2$\times$2$\times$2 containing 160 atoms are chosen. The 2nd- and 3rd-force constants are extracted from the temperature dependent effective potential (TDEP)\cite{Hellman2013}, which represent the phonon-phonon interactions and electron-phonon interactions at current temperatures as accurate as possible with a finite order. The results of \textit{ab initio} molecular dynamics simulations with time interval of 1 fs and total simulation steps of 4000 are set as the training data of DeePMD-kit\cite{Wang2018b} to train for the machine learning potential (MTP). Then, the obtained MTP file will be the input file for the LAMMPS package\cite{Thompson2022}. At different temperatures, the canonical (NVT) ensemble was first employed for running 1 ns with a time step of 1 fs to bring the system to thermal equilibrium. Then the system was sampled by 0.1 ps within a period time of 100 ps. These samples were used as the inputs of TDEP to compute the 2nd(3rd)-order IFCs. The truncation radii for 2nd(3rd)-order IFCs are set to be 6.94$\rm{\r{A}}$ and 13.43$\rm{\r{A}}$, respectively.

\section*{Conflicts of interest}
There are no conflicts to declare.

\section*{Acknowledgements}

This work is supported by Natural Science Foundation of China (12304038), Huzhou Natural Science Foundation (2023YZ50), Shanghai Municipal Natural Science Foundation under Grant (19ZR1402900), Sichuan Natural Science Foundation (2022NSFSC1193), the Startup funds of Outstanding Talents of UESTC (A1098531023601205), National Youth Talents Plan of China (G05QNQR049).


\section*{Reference}
\providecommand{\latin}[1]{#1}
\makeatletter
\providecommand{\doi}
  {\begingroup\let\do\@makeother\dospecials
  \catcode`\{=1 \catcode`\}=2 \doi@aux}
\providecommand{\doi@aux}[1]{\endgroup\texttt{#1}}
\makeatother
\providecommand*\mcitethebibliography{\thebibliography}
\csname @ifundefined\endcsname{endmcitethebibliography}
  {\let\endmcitethebibliography\endthebibliography}{}

\end{document}